\documentclass[twocolumn,showpacs,preprintnumbers,amsmath,amssymb]{revtex4}
\usepackage{graphicx}% Include figure files
\usepackage{dcolumn}% Align table columns on decimal point
\usepackage{bm}% bold math

\begin{document}

\title{
Spontaneous Collapse of Unstable Quantum Superposition State: \\
A Single-Particle Model 
of Modified Quantum Dynamics
}
%\title{
%Modified Quantum Dynamics with Spontaneous Wavefunction Collapse:\\
%A Single-Particle Model
%}

\author{Takuya Okabe}
 \email{ttokabe@ipc.shizuoka.ac.jp}
\affiliation{
Faculty of Engineering, Shizuoka University, 
Hamamatsu 432-8561, Japan
}

\date{\today}

\begin{abstract}
We propose
a modified dynamics of quantum mechanics,
in which 
classical mechanics of a point mass
derives intrinsically
in a massive limit
of a single-particle model.
On the premise that a position basis plays a special role in
 wavefunction collapse,
we deduce to formalize 
spontaneous localization of wavefunction
on the analogy
drawn from thermodynamics, in which
a characteristic energy scale and a time scale are introduced
 to separate quantum and classical regimes.
\end{abstract}

%03.65.Ta 	Foundations of quantum mechanics; measurement theory
%02.50.Ey 	Stochastic processes 
%07.05.Tp 	Computer modeling and simulation


\maketitle
There is a significant recent increase of interest 
in the foundations of quantum mechanics (QM),
owing to the technological progress
particularly in the field of quantum optics
to investigate
%in experimental investigation of 
individual stochastic behaviors of quantum systems.
%which
%%Among others, this 
%is particularly
%illuminated by the remarkable development 
%in the field of quantum optics.
%In particularl,
%the progress of technology in the field of quamtum optics
%is remarkable. 
%
Nevertheless, it would be fair to state that
conceptual and philosophical difficulties of QM are still left behind
persistently.
Indeed, 
we are so accustomed to the classical notion of definite position 
in real life 
that we can scarcely imagine
what a superposition of 
macroscopic states localized at different positions
would look like.
On the one hand, there are many who advocate that
such a problematic state must be excluded 
in terms of a proper interpretation adapted to within 
the present formalism of QM.
On the other hand, 
there are researchers questing for 
the clearcut solution 
of the so-called ``measurement'' problem
on the basis of a realistic viewpoint,
by modifying the dynamical formalism of QM
\cite{rf:hazy, rf:gisin, rf:diosi,rf:pearle,rf:grw,
rf:percival,rf:hughston,rf:ad}.
From the latter viewpoint,
the aim of this paper is to make 
such an attempt
to formalize wavefunction collapse
without being involved in 
many-particle features of a macroscopic body,
so that the conceptual gap between classical and quantum mechanics
may be bridged mathematically
within a single-particle formalism.

Ghirardi, Rimini and Weber (GRW)
made a breakthrough with a many-particle model 
postulating {discontinuous} collapse\cite{rf:grw},
which was soon developed into a continuous model by
Pearle\cite{rf:pearle}.
Our proposal presented below has similar physical implications
as the original discontinuous GRW model,
where also two parameters are introduced.
In the GRW model, a many-particle wavefunction 
is subjected to spontaneous localizations
in a multidimensional configuration space of the micro-constituents, 
of which the frequency and the localization function,
or the localization length,
are postulated to be universal,
$\lambda \simeq 10^{-16}$sec$^{-1}$ and 
$a\simeq 10^{-5}$cm,
respectively.
The GRW localization mechanism is such that 
the collapse rate effectively
increases as the number $N$ of the constituents
increases,
so that a macroscopic object comprising
an Avogadro number of constituents 
collapses extremely rapidly, 
at a rate of
$N\lambda \simeq 10^7$sec$^{-1}$.
In contrast, 
we shall postulate 
the universal collapse rate $\gamma_0=\tau_0^{-1}$ independently of
system size.
On the other side, instead of positing the universal length,
we scheme that
the localization length scale for a given particle state
%width of a center-of-mass coordinate
is variationally determined case by case
by the mass $m$ of the particle, an external potential, and
another constant $T_0$ of the dimensions of energy.
In this respect, 
our proposal may seem more complicated than the GRW model.
However, we argue that the complication is compensated 
by our simple premise 
that, unlike the previous 
works\cite{rf:diosi,rf:pearle,rf:grw,rf:percival,rf:hughston,rf:ad},
we intend to cope with the measurement problem 
without resort to many-particle treatment. 
In effect, we 
regard the problem essentially
{as a single-particle problem} of the modified dynamics.
This must be the simplest and non-trivial option to attack the problem,
which however seems not to be undertaken thoroughly so far.
Specifically, 
in essence,
we shall attach no fundamental relevance to 
the number $N$ of particles comprising a system.
Therefore, 
we focus directly and solely on the effective single-particle model
%(\ref{model}) 
proper, of which we still stress
the wide potential applicability 
to offset 
the primitive status of the present proposal 
compared with the 
highly developed
many-particle 
treatments\cite{rf:diosi,rf:pearle,rf:grw,rf:percival,rf:hughston,rf:ad}.
Though we are motivated 
to provide a plain scheme 
to derive classical mechanics
in a large mass limit $m\rightarrow \infty$ of the
model, one should bear in mind that
our modified dynamics may predict non-trivial
``classical'' behaviors for
an extremely weakly coupled microscopic system, like
a quantum particle trapped in a deep double well potential.

To begin with, 
the special status we assign 
below (in (\ref{S'}))
to the position variable ${\bf r}$ 
of wavefunction $\Psi({\bf r})$ is apparent from 
its role in the classical limit.
According to Ehrenfest's theorem, the Newton equation 
\begin{equation}
 m \frac{{\rm d}^2 \bar{\bf r}}{{\rm d}t^2} =-\nabla V(\bar{\bf r})
\label{newton} 
\end{equation}
for the point mass
$\bar{\bf r} \equiv \langle {\bf r} \rangle$
derives from the Schr\"odinger equation
\begin{eqnarray}
{\rm i}{\hbar}\frac{\partial \Psi}{\partial t}
=\hat{H}\Psi
=
\left(-\frac{\hbar^2 \nabla^2}{2m} +V({\bf r})
\right)\Psi,
\label{Schrodingereq} 
\end{eqnarray}
in the limit $(\delta {\bf r})^2=\langle {\bf r}^2\rangle-
\bar{\bf r}^2\rightarrow 0$.
However,
even if the limit should hold true at a moment,
it cannot remain so
for ever
according to (\ref{Schrodingereq}),
or we cannot justify 
the general validity of (\ref{newton}) on the basis of 
(\ref{Schrodingereq}) alone.
Hence, we stipulate
a modified intrinsic dynamics of $\Psi({\bf r})$ to
keep $|\delta {\bf r}|$ finite so that classical mechanics 
(\ref{newton}) of the mass point $\bar{\bf r}(t)$ 
derives approximately and holds for good.
Here, as a premise,
we do not resort to 
any other system, degrees of freedom, and whatever
cause but collapse to establish (\ref{newton}).
It is stressed that 
%we are primarily interested in 
%the {\it individual}
%
%(instead of statistically averaged)
%behavior of a particular system.
a matter of our interest is 
to follow time development of 
a particular system individually (instead of statistically).

In order to describe a mode of wavefunction collapse, 
we introduce a set of $N$ real functions
$P_n({\bf r})$ ($n=1,2,\cdots, N$)
corresponding to the $N$ potential results of collapse.
Given the normalized wavefunction $\Psi_t({\bf r})$
of a particle at time $t$, 
\begin{equation}
 \int {\rm d}{\bf r}\rho({\bf r})=1, \qquad 
\rho({\bf r})=|\Psi_t({\bf r})|^2,
\label{int drrho=1} 
\end{equation}
our idea is to fix the localization functions
$P_n({\bf r})$ so as to minimize 
the spatial extension $(\delta {\bf r})^2$ 
of the ensemble $\Phi_n({\bf r}) \propto P_n \Psi$
realized by the spontaneous collapse.
There are various ways to express this mathematically.
Postulating that
the collapse outcomes are given by
\begin{equation}
\Phi_n=\frac{1}{\sqrt{w_n}}P_n \Psi_t,
\label{2ndlaw-Phin}
\end{equation}
where
\begin{equation}
 w_n=\int{\rm d}{\bf r} \rho({\bf r}) P_n({\bf r})^2,
\label{w_n}
\end{equation}
we propose that 
$P_n({\bf r})$ are determined 
so as to maximize 
\begin{equation}
 S'= \sum_n w_n \int |\Phi_n|^2
(\log |\Phi_n|^2 -1 ) {\rm d}{\bf r},
\label{S'}
\end{equation}
under the constraint 
\begin{equation}
T_0 \Delta S= \Delta E,
\label{T0DeltaS>DeltaE}
\end{equation}
where $\Delta S$ and $\Delta E$
represent the physical changes of entropy and energy 
accompanied by the collapse $\Psi\rightarrow \Phi_n$,
which, from our standpoint, is regarded as a real process of nature.
In (\ref{T0DeltaS>DeltaE}), 
we introduced the model parameter $T_0$ of the dimensions of energy.
The functions $P_n$ should be
%in order for (\ref{w_n}) to be regarded as the collapse probability,
fixed variationally under the constraint 
\begin{equation}
 \sum_n P_n({\bf r})^2=1,\qquad  0\le P_n({\bf r})\le 1.
\label{sumnPnP2=1}
\end{equation}
Thus we divide the whole space ${\bf r}$ into 
numbers of patches indexed by $n$, 
in one of which the particle is found 
as a result of spontaneous collapse.
%In a sense, (\ref{sumnPnP2=1}) is a decomposition of uniy.
The entropy $S=-{\rm Tr} (\hat{\rho} \log \hat{\rho})$
and the energy $E={\rm Tr} (\hat{\rho} \hat{H})$
to be used for (\ref{T0DeltaS>DeltaE})
are customarily defined in terms of 
the density matrix $\hat{\rho}$ of the collapse outcomes,
\begin{equation}
 \langle {\bf r}|\hat{\rho}|{\bf r'}\rangle
=\sum_n w_n \Phi_n({\bf r})^*\Phi_n({\bf r}').
%=\Psi({\bf r})^*\Psi({\bf r}')\sum_n P_n({\bf r})P_n({\bf r}')
\label{densitymatrix} 
\end{equation}
In particular, 
for $\hat{H}$ defined in (\ref{Schrodingereq}), we obtain
\begin{equation}
\Delta E=
\frac{\hbar^2}{2m}
\int {\rm d}{\bf  r} \rho({\bf r})
\sum_n 
\left(\nabla P_n({\bf r})
\right)^2 >0.
\label{DeltaE=frachbar2}
\end{equation}

%
%\vspace*{5cm}
%
%in terms of $P_n({\bf r})$
%
%\begin{eqnarray}
%(\delta {\bf r})^2=
%\sum_n w_n 
%\left(
%{\langle \Phi_n| {\bf r}\cdot {\bf r}|\Phi_n\rangle} 
%-|{\langle \Phi_n| {\bf r}|\Phi_n\rangle}|^2 
%\right)
%\nonumber
%%&=&
%%{\langle \Psi| {\bf r}\cdot {\bf r}|\Psi\rangle} 
%%-
%%\sum_n \frac{
%%|\int {\rm d}{\bf r} \rho({\bf r}) P_n({\bf r})^2 {\bf r}|^2 }
%%{\int {\rm d}{\bf r}\rho({\bf r}) P_n({\bf r})^2},
%%\nonumber
%\end{eqnarray}
%which is
%the spatial variance of the statistical ensemble $\{\Phi_n\}$
%that would be realized by the collapse.
%
%
%%kokomade
%
%maximize
%
%\begin{eqnarray}
%\sum_n \frac{
%\left|
%\displaystyle \int {\rm d}{\bf r} \rho({\bf r}) P_n({\bf r})^2 
%{\bf r}
%\right|^2 }
%{\displaystyle \int {\rm d}{\bf r}\rho({\bf r}) P_n({\bf r})^2}
%\label{maximize}
%\end{eqnarray}
%under the conditions
%and
%
%where 
%\begin{eqnarray}
% \Delta S&=&-\sum_n w_n \log w_n,
%\label{DeltaS}
%\\
%\end{eqnarray}
%and
%\begin{eqnarray}
% \Delta E&=&
%\frac{\hbar^2}{2m}
%\int {\rm d}{\bf  r} \rho({\bf r})
%\sum_n 
%\left(\nabla P_n({\bf r})
%\right)^2,
%\label{DeltaE}
%\end{eqnarray}
%in which $\hbar$ is the Planck constant, and
%$m$ is the mass of the particle.
%

In terms of $\{P_n\}$, 
and a constant $\gamma_0$ of the dimensions of
frequency, 
we postulate the following modified dynamics 
to relate $\Psi_{t+\Delta t}({\bf r})$ with $\Psi_t({\bf r})$,
where $\Delta t$ is an infinitesimal time interval.
%From the two alternative evolutions, deterministic or stochastic,
The continuous deterministic evolution
(i) applies only with probability $1-\gamma_0 \Delta t$,
while the discontinuous stochastic evolution 
(ii) or (ii') applies with probability $\gamma_0 \Delta t$:
(i) The Schr\"odinger time evolution:
\begin{equation}
 \Psi_{t+\Delta t}
=
\left(
1-\frac{{\rm i}}{\hbar}\hat{H}\Delta t
\right)
 \Psi_{t}.
\label{1stlaw}
\end{equation}
%where the Hamiltonian is
%\begin{equation}
% H=-\frac{\hbar^2 \nabla^2}{2m} +V({\bf r}).
%\label{model} 
%\end{equation}
(ii) Stochastic collapse:
If $\{P_n\}$ is uniquely determined,
then
the $n$-th result 
from among the $N$ possibilities (\ref{2ndlaw-Phin})
is realized 
with the probability $w_n$
(the absolute probability $w_n\gamma_0\Delta t$),
\begin{equation}
 \Psi_{t+\Delta t}
=\Phi_n.
%\equiv
%\frac{
%P_n \Psi_t}{\sqrt{w_n}}. 
\label{2ndlaw}
\end{equation}
(ii') If the set $\{P_n\}$ is not uniquely determined from the given
$\rho({\bf r})$,
one set is chosen at random with equal a priori probability,
then apply (ii).

We have the two parameters 
$T_0$ and $\tau_0\equiv \gamma_0^{-1}$
to characterize the present collapse model.
%
%In the above, 
%$T_0$ and $\tau_0\equiv \gamma_0^{-1}$ are 
%the two parameters of the model.
%postulated to be universal
%constants independent of $m$ and 
%the potential energy $V({\bf r})$.
One can take the limit $\Delta t\rightarrow 0$ without changing the
physical consequences, viz.,
the temporal sequence of $\Psi_t ({\bf r})$.
Note that the probability to keep following (i)
during $N$ steps of $\Delta t=\tau/N$ is given by
\(
\left(1-\frac{\gamma_0 \tau}{N}\right)^N \rightarrow 
\exp\left(-\gamma_0 \tau
\right)
\) as $N\rightarrow \infty$.
Accordingly,
the quantum state $\Psi({\bf r})$ is always
subjected to 
spontaneous collapse with the constant decay rate $\gamma_0$.
In particular, 
we recover (\ref{Schrodingereq})
in the limit $\gamma_0\rightarrow 0$,
irrespective of $T_0$.
It may be of interest to note the methodological
similarity of the present dual dynamics (i) and (ii)
with the so-called Monte Carlo wavefunction approach 
to solve the optical Bloch equations\cite{rf:dcm,rf:hjc},
in which 
the decay rate of an unstable state plays
the counterpart of our $\gamma_0$.
The latter approach follows the conventional procedure
to invoke ``measurement'' to realize collapse.

%Let us remark the physical significance of the above postulates.
The probability $w_n$ defined in (\ref{w_n}) adds up to unity,
\(
 \sum_n w_n=1, 
\)
owing to 
(\ref{sumnPnP2=1}) and (\ref{int drrho=1}),
while
the collapse outcomes $\Phi_n({\bf r})$ are normalized
by $w_n$ in (\ref{2ndlaw-Phin}),
\(
 \int{\rm d}{\bf r} |\Phi_n({\bf r})|^2=1.
\)
In general, however, the outcomes $\Phi_n$ are not orthogonal,
%we observe the non-orthogonality of
%the $\{\Phi_n\}$, 
or $P_n({\bf r}) P_m({\bf r})\ne 0$ for $n\ne m$.
In the special case of 
the strict localization by 
the step function $P_n({\bf r})^2=P_n({\bf r})$, for which
$P_n({\bf r})=0$ or 1,
the postulate (\ref{w_n}) reproduces
the conventional Born formula,
\(
w_n =\left|
\langle\Phi_n| \Psi\rangle
\right|^2
=\int_{P_n({\bf r})=1} {\rm d}{\bf r} |\Psi({\bf r})|^2.
\)
Consequently, 
Born's probability rule is reproduced approximately
when we find
$P_n({\bf r}) P_m({\bf r})\simeq 0$ for $n\ne m$
(cf. below (\ref{Psirsumncn})).

%To maximize (\ref{maximize}) means to minimize 
%\begin{eqnarray}
%(\delta {\bf r})^2&=&
%\sum_n w_n 
%\left(
%{\langle \Phi_n| {\bf r}\cdot {\bf r}|\Phi_n\rangle} 
%-|{\langle \Phi_n| {\bf r}|\Phi_n\rangle}|^2 
%\right)
%\nonumber\\
%&=&
%{\langle \Psi| {\bf r}\cdot {\bf r}|\Psi\rangle} 
%-
%\sum_n \frac{
%|\int {\rm d}{\bf r} \rho({\bf r}) P_n({\bf r})^2 {\bf r}|^2 }
%{\int {\rm d}{\bf r}\rho({\bf r}) P_n({\bf r})^2},
%\nonumber
%\end{eqnarray}
%which is
%the spatial variance of the statistical ensemble $\{\Phi_n\}$
%that would be realized by the collapse.

It is always possible to find a non-trivial set $\{P_n\}$ 
for arbitrary $\rho({\bf r})$.
To give a simple example, 
one may assume the trial functions
\begin{equation}
 P_\pm ({\bf r})^2 =
\frac{1}{2}\left(1\pm 
\tanh 
\left[
({{\bf r}\cdot {\bf i}-X})/{l}
\right]
\right)
\label{tanh}
\end{equation}
%and $P_2({\bf r})^2=1-P_1({\bf r})^2$
to satisfy (\ref{sumnPnP2=1}),
fix the direction of the unit vector ${\bf i}$,
and vary the parameters $X$ and $l$ 
to maximize $S'\simeq \log |\delta {\bf r}|^{-3}$ 
%(or to minimize $(\delta {\bf r})^2$) 
under the condition 
(\ref{T0DeltaS>DeltaE}).
%
%where the inequality may be replaced with
%equality in practice.
Thus, in principle, there is no problem to
implement our modified dynamics
for numerical use in place of the collapse-free Schr\"odinger equation
(\ref{Schrodingereq}).
With respect to the case (ii'),
we have in mind the special situations where symmetry possessed by 
$\hat{H}$ and $\rho({\bf r})$ is spontaneously broken by collapse,
or by a specific choice of $\{P_n({\bf r})\}$.
This holds, for example, 
in the reduction of a spherically symmetric outgoing wave $\Psi(|{\bf r}|)$
under the spherically symmetric potential $V(|{\bf r}|)$.

\begin{figure}[t]
\begin{center}
\includegraphics[width=0.33\textwidth]{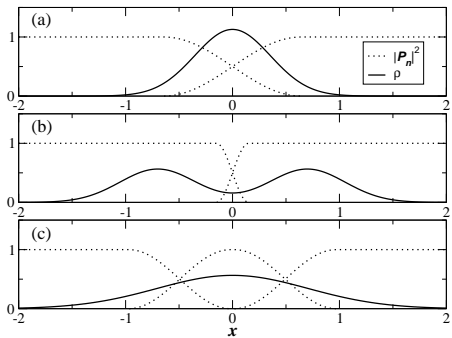}
\caption{
Schematic examples of $\rho(x)$ 
(solid lines), and $P_n(x)^2$ (dotted lines)
for $\hbar=T_0=m=1$.
%According to (\ref{T0DeltaS>DeltaE}),
Extended
wavepackets are spontaneously reduced down to 
finite widths of order 
$\lambda_0/2$ ($\simeq 1$) at most.
}\label{fig:rhoandPn}
\end{center}
\end{figure}

%To show that 
The above model is not so artificial as it may first look.
In fact, 
%the above postulates,
%On the one hand, 
%(\ref{DeltaS}) represents the change of entropy, 
%while (\ref{DeltaE}) is the change of energy
%accompanied in the spontaneous irreversible process (ii),
%\begin{eqnarray}
% \Delta E&=&\sum_n w_n 
%\langle \Phi_n| H|\Phi_n\rangle
%-
%\langle \Psi| H|\Psi\rangle
%\nonumber\\
%&=&
%-\frac{\hbar^2}{2m}\int {\rm d}{\bf r}
%\left(
%\sum_n w_n \Phi_n^* \nabla^2 \Phi_n
%-  \Psi^* \nabla^2 \Psi
%\right),
%\nonumber
%\end{eqnarray}
%which equals (\ref{DeltaE})
%because of
%\(
% \sum_n P_n({\bf r}) \nabla P_n ({\bf r})=0  
%\)
%and 
%\(
% \sum_n\left(\nabla P_n({\bf r})
%\right)^2+
% \sum_n P_n({\bf r}) \nabla^2 P_n ({\bf r})=0
%\)
%due to (\ref{sumnPnP2=1}).
%Therefore, 
the condition (\ref{T0DeltaS>DeltaE}) as well as (ii') 
suggest themselves naturally
on the analogy of the second law of thermodynamics.
It is the established fact that
a thermodynamic state immersed in the heat bath at temperature $T_0$ 
suffers the spontaneous irreversible change
under the condition $T_0\Delta S\ge \Delta E$.
%(\ref{T0DeltaS>DeltaE}) is satisfied.
Although there is no logical necessity that
the law governing the microscopic irreversible process of our concern,
wavefunction collapse,
must bear resemblance to that of thermodynamics,
we argue that the proposed postulates would be justified 
if only by relying on the analogy, and 
the consequences must be worth close inspection.
In short, 
the postulates are physically interpreted as follows:
Wavefunction collapse tends to reduce the spatial 
extent $(\delta {\bf r})^2$ of
the wavepacket at intervals (spontaneous localization\cite{rf:grw}), 
while respecting the ``thermodynamic''
constraint (\ref{T0DeltaS>DeltaE}).
To help understand how (ii) works,
some examples of $\rho(x)$ and $P_n(x)^2$ 
are schematically displayed in Fig.~\ref{fig:rhoandPn}.

Without detailed calculation, 
we can
make an order estimate of how small $(\delta {\bf r})^2$ is reduced.
According to (\ref{T0DeltaS>DeltaE}),
even in the absence of the external potential $V({\bf r})=0$,
the energy scale $T_0$ introduces a length scale 
\begin{equation}
\lambda_0\equiv 
\hbar\sqrt{\frac{2\pi}{mT_0}},
\label{lam0}
\end{equation}
which is nothing but 
the thermal de Broglie wavelength at ``temperature'' $T_0$.
Accordingly, we obtain the finite wavepacket width of order
$|\delta {\bf r}| \simeq \lambda_0 \propto m^{-1/2}$,
which brings about our purpose to reproduce classical mechanics 
of the point mass ${\bf r}(t)$
in the limit $\lambda_0\rightarrow 0$, or $m\rightarrow \infty$.
The finite wavepacket width (\ref{lam0}) from (\ref{T0DeltaS>DeltaE}) 
is just as expected 
by the thermodynamic analogy
mentioned above.
In effect, the degree of localization $(\delta {\bf r})^2$
depends on
the energy $\Delta E>0$
required for the localization.
If it were not for the constraint (\ref{T0DeltaS>DeltaE}),
or if we let $T_0\rightarrow \infty$,
% to replace (\ref{T0DeltaS>DeltaE}) with $\Delta S>0$,
the hypothetical spontaneous collapse would reduce
$(\delta {\bf r})^2$ without limit,
and the established results 
of quantum mechanics at a short length scale 
must be spoiled altogether.
Thus, in the present model, 
the 
crossover length scale presumed between 
classical and quantum regimes
is regulated by $T_0$ in a controlled and universal manner.

\begin{figure}[t]
\includegraphics[width=0.47\textwidth]{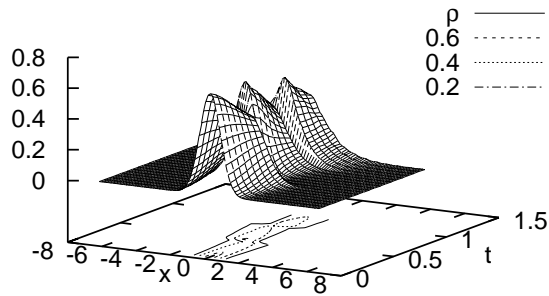}
\caption{
\label{fig:free2}
A schematic figure of 
the temporal evolution of a free wavepacket density
 $\rho(x,t)$ expected in one dimension.
%Two stochastic collapses are recognized.
Wavefunction collapse with the frequency $\gamma_0=\tau_0^{-1} (\simeq 1)$
keeps $|\delta x|$ from diffusing.
In the limit (\ref{T0tau0llhbar}), 
the time scale $\sim \hbar/T_0$ of diffusion is 
slower than the collapse interval $\sim \tau_0$.
}
\end{figure}

To put it concretely, in order to guarantee 
$\lambda_0 > 10$m for electron
while $\lambda_0 < 1\mu$m for a tiny particle of a nanogram
(either too large or small for the crossover regime
to be detected unexpectedly),
we estimate
\begin{equation}
10^{-49}\ {\rm J} <T_0< 10^{-39}\ {\rm J}. 
\label{estimateT0}
\end{equation}
Without being affected by the collapse,
the wavepacket evolves according to (\ref{1stlaw})
for the duration of order $\tau_0$.
In the meantime,
the width $(\delta {\bf r})^2\simeq \lambda_0^2$
grows up to $\lambda_0^2\left(1+ 
(\frac{T_0 \tau_0}{4\pi \hbar})^2)\right)$,
so that the second length scale must be introduced if
$T_0\tau_0\gg 4\pi \hbar$.
However, with the above estimate taken for granted, 
this limit must be rejected physically,
because we hardly accept $\tau_0$ as long as 
$\hbar/T_0 > 10^5$~second on physical grounds.
Hence we assume the limit
\begin{equation}
T_0\tau_0\ll 4\pi \hbar
\label{T0tau0llhbar} 
\end{equation}
without invalidating the model, 
and consequently, the single length scale (\ref{lam0}) in free space.
Specifically, the collapse time $\tau_0$ must be 
longer than characteristic time scales of microscopic phenomena
because the postulated collapse must more or less 
affect conventional quantum results following from
the Schr\"odinger equation.
At the same time, $\tau_0$ must be
short enough to ensure that 
we have never observe such 
a problematical superposition of 
the ``classical'' states with ``distinct'' spatial configurations.
Therefore, physically, it is the vital point of the model 
that experiments do not exclude 
%a sensible choice  $\tau_0$.
%the possibility for 
the non-trivial scale $\tau_0$.
%Needless to say,
%we should assess the distinctness of position
%$\bar{\bf r}$ in comparison with
%$|\delta {\bf r}|$.
By way of illustration, 
the density profile $\rho(x,t)$ of a free wavepacket
expected in the modified dynamics is 
schematically displayed in Fig.~\ref{fig:free2}.
%In principle, the finite $|\delta {\bf r}|$ predicts
%the non-trivial quantum corrections to (\ref{newton}).

Let us discuss a particle under the external potential,
where 
another length scale 
is brought about by the potential $V({\bf r})$.
In the limit $m\rightarrow 0$ where
$\lambda_0$ exceeds the length scale,
the non-trivial effects depending on $T_0$ become negligible.
This must be the case for 
well-established quantum mechanical
phenomena of microscopic systems where
wavefunction collapse is not apparently resorted to for their explanation.
For example, for the atom with $V({\bf r})=-e^2/r$, 
we have the ground state with
$|\delta {\bf r}| \simeq a_{\rm B}=\hbar^2/me^2$,
the Bohr radius.
It is straightforward to check that
the collapse (ii) will not affect the bound state $\Psi({\bf r})$
in
 the limit 
$|\delta {\bf r}| \simeq a_{\rm B}\ll \lambda_0$.
This is 
because $P_n({\bf r})$ then
turn out to be almost constant over the short 
length scale of order $a_{\rm B}$,
so that
all the right-hand sides of 
(\ref{1stlaw}) and (\ref{2ndlaw}),
the possible states at $t+\Delta t$, become 
physically
indistinguishable
from each other
($\Delta E=T_0\Delta S\simeq 0$).
In other words, 
in this quantum limit, 
the characteristic energy scale (\ref{estimateT0})
% of collapse
to cause non-standard change of state
%to cause non-trivial effects
is too small to be 
compared with other standard
perturbations of practical relevance.

On the contrary, 
if $\Psi({\bf r})$ is represented by a superposition of localized
wavepackets, 
comprising several distinct peaks of amplitude
well separated from each other, 
it becomes possible to reduce $\Psi({\bf r})$ to
the ensemble of single-peaked localized states $\Phi_n({\bf r})$.
To put it more precisely, 
as we predict the localization to the degree 
to cost $\Delta E \simeq T_0$, 
the effect, though minute energetically,
can happen to be quite conspicuous as the case may be.
We argue that 
this is relevant in
situations encountered in position measurement
of a quantum particle, where
%In measurement,
a superposition state 
of localized wavepackets realized 
through interaction with detectors
is prepared according to (i)
within a microscopic time scale shorter than $\tau_0$.

For example, consider the ``pre-measurement'' superposition 
\begin{equation}
% \Psi ({\bf r})=\sum_n c_n \phi_n({\bf r}),
\Psi({\bf r})=c_+ \phi_+ + c_- \phi_-, 
%\qquad\phi_\pm=\phi_\sigma({\bf r}\mp {\bf R}),
\label{Psirsumncn}
\end{equation}
%$\Psi({\bf r})=c_+ \phi_+ + c_- \phi_-$
of the localized wavepacket $\phi_\sigma({\bf r})=
{{\rm e}^{-|{\bf r}|^2/2\sigma^2}}/{(2\pi \sigma^2)^{3/2}}$,
where $\phi_\pm=\phi_\sigma({\bf r}\mp {\bf R})$ 
and $|c_+|^2+|c_-|^2=1$.
%
%\begin{equation}
% \Psi({\bf r})=c_+ \phi_\sigma({\bf r}-{\bf R})
%+c_- \phi_\sigma({\bf r}+{\bf R}),
%\label{psi=c+phi+c-}
%\end{equation}
%where $|c_+|^2+|c_-|^2=1$.
As soon as the separation between the two wavepackets exceeds 
their own width, $|{\bf R}|\gg \sigma$,
the state is localized spontaneously
%one will obtain one of the localized states
$\Psi\rightarrow \frac{c_n}{|c_n|}\phi_n$ with the probability 
$w_n\simeq |c_n|^2$
within the lifetime $\tau_0$.
%according to the modified dynamics.
To put it concretely, 
with (\ref{tanh}),
we find ${\bf i}={\bf R}/|{\bf R}|$, $X=0$ and
$l\simeq \lambda_0^2 {\rm e}^{-|{\bf R}|^2/\sigma^2}/\sigma\ll \sigma$,
%The ``pre-measurement'' superposition state would be represented as
%\begin{equation}
% \Psi ({\bf r})=\sum_n c_n \phi_n({\bf r}),
%\label{Psirsumncn}
%\end{equation}
%where $\phi_n({\bf r})$ are normalized,
%compactly localized around 
%different points ${\bf R}_n$, and they are spatially distinct
%so that
%$|\langle \phi_m |\phi_n\rangle| \simeq 0$ 
%for $m\ne n$.
%Then, 
so that $P_\pm$ 
singles out $\phi_\pm$ approximately,
$P_n\phi_m\simeq \phi_n \delta_{mn}$
and $P_+ P_-=0$.
%and obtain
%one of the localized states
%$\Psi\rightarrow \frac{c_n}{|c_n|}\phi_n$ with the probability 
%$w_n\simeq |c_n|^2$
%within the time $\tau_0$.
In this context, 
$\tau_0$ is interpreted as
the time scale for the ``measurement'' to complete. 
%or for a definite result on the pointer basis ${\bf R}_n$
%to determine instantly.
Thus
 we envisage the quantum jump in measurement
as
a single-particle phenomenon of the system
to be measured, in which the role of the apparatus is 
to provide the external potential $V({\bf r},t)$ 
devised to realize such a superposition state
like (\ref{Psirsumncn}).
In this model,
the amplification processes which would generally follow
to develop the definite result $\phi_n$
macroscopically
are of secondary importance.

Incidentally, it is also of interest to consider
the opposite ``adiabatic'' limit that 
$V({\bf r},t)$ changes so gradually that
(\ref{Schrodingereq}) would not change
$\rho({\bf r},t)$ appreciably within $\tau_0$.
Then, despite the collapse,
$(\delta {\bf r})^2$ changes almost continuously.
Therefore, 
according to the modified dynamics,
$\rho({\bf r},t)$
will evolve along a ``classical'' trajectory
as in Brownian motion (cf. Fig.~\ref{fig:free2}),
without going through 
such an abrupt change suffered by (\ref{Psirsumncn}).
%Note a key role of $\rho({\bf r},t)=|\Psi({\bf r},t)|^2$
%in our model.
% from (\ref{maximize}) to (\ref{DeltaE}).
To illustrate the above schematic argument more concretely,
numerical as well as analytical results for simple cases of
one-dimensional systems,
as shown in Figs.~\ref{fig:rhoandPn} and \ref{fig:free2}
which are indeed our preliminary results,
will be presented in future work\cite{rf:okabe}.

In conclusion,
starting from a single-particle problem of
a GRW-like modified Schr\"odinger dynamics
devised to derive classical mechanics of 
the point particle $\bar{\bf r}(t)$,
we ended up with a single-particle picture of measurement.
We introduced
the constant energy $T_0$, 
in place of the constant length $a$ of the GRW model,
and the universal collapse time $\tau_0$,
both of which are to be fixed experimentally,
as are the parameters 
in the other collapse 
models\cite{rf:diosi,rf:pearle,rf:grw,rf:percival,rf:hughston,rf:ad}.
The model is based on 
the ideas of spontaneous localization\cite{rf:grw}
and the second law of thermodynamics.

The author would like to thank Prof. S. Takagi for stimulating
discussions.

%$Last Update: Wed Oct 13 14:24:09 2004 $

\end{document}